\documentclass[aps, prb, reprint, showpacs]{revtex4-1}

\usepackage{graphicx}
\usepackage{dcolumn}
\usepackage{bm}
\usepackage{wrapfig}
\usepackage{scalefnt}

\usepackage{array,booktabs,tabularx,tabulary}
\usepackage{lipsum}
\newcolumntype{L}{>{\raggedright\arraybackslash}X}
      \usepackage{siunitx}
      \usepackage{caption}
\usepackage[export]{adjustbox}
\usepackage[colorlinks=true, linkcolor=red, citecolor=blue, urlcolor=blue, linktoc=page, bookmarks=false, pdfstartview={FitH}, pdfborder={0 0 0.0 [3 3]}]{hyperref}

\usepackage{amsmath}
\usepackage{natbib}
\usepackage{natmove}
\usepackage{dcolumn}
\newcolumntype{.}{D{.}{.}{-1}}

\def\9{\left(}
\def\0{\right)}

\def\braket#1{\mathinner{\left\langle{#1}\right\rangle}}
\begin{document}
\title{Electronic, magnetic and ferroelectric properties of  rhombohedral AgFeO$_2$: an {\em {ab initio} } study}
\author{Jayita Chakraborty}
\altaffiliation{Present address: Department of Physics, Indian Institute of Science Education and Research Bhopal, Bhauri, Bhopal 462066, India}

\author{Indra Dasgupta}
\affiliation{Department of Solid State Physics, Indian Association for the Cultivation of Science, Jadavpur, Kolkata 700032, India}
\date{\today}

\begin{abstract}
Using first principle calculations under the framework of density functional theory we have investigated the electronic structure, magnetism and ferroelectric polarization in the triangular lattice antiferromagnet AgFeO$_2$, and its comparison to the isostructural system CuFeO$_2$. Our calculations reveal that spin orbit interaction plays an important role in determining the magnetic property of AgFeO$_2$ and is possibly responsible for its different magnetic ground state in comparison to CuFeO$_2$. Calculations of ferroelectric polarization of AgFeO$_2$ suggest that the spontaneous polarization arises from noncollinear spin arrangement via spin-orbit coupling. Our calculations also indicate that in addition to electronic contribution, the lattice mediated contribution to the polarization are also important for AgFeO$_2$.  
\end{abstract}
\maketitle

In recent times two dimensional  triangular lattice antiferromagnets have attracted much attention both theoretically as well as experimentally because of the fascinating magnetic properties displayed by them due to geometric frustration.\cite{Balents2010, LawesPRL2004, kotikagome} In addition, some of these compounds exhibit ferroelectricity. ABO$_2$-type compounds with delafossite structure provide a good example of triangular lattice antiferromagnets (TLA) and present an opportunity to study the influence of geometric spin frustration in magnetic properties.\cite{Kimuraprl2009, Zhang2011} In ABO$_2$ compounds, the A-site cation has completely filled $d$ orbitals, $\9\text{Cu}^+\text{-}3d^{10} ~\text{and}~\text{Ag}^+\text{-}4d^{10}\0$, while the B-site cation has partially filled $d$ orbitals $\9 \text{Cr}^{3+}\text{-}3d^{3}~ \text{and}~ \text{Fe}^{3+}\text{-}3d^{5}\0$. 
Examples of ABO$_2$  systems include CuFeO$_2$, AgCrO$_2$, CuCrO$_2$, AgFeO$_2$ etc.\cite{Seki2008,Vasiliev2010,Malvestuto2011} 
The magnetic ground state of the delafossite CuFeO$_2$ has $\uparrow\uparrow\downarrow\downarrow$ collinear spin structure with their spins parallel to the $c$ axis.\cite{MitsudaJPSJ1991} It shows multistep metamagnetic phase transitions when a varied magnetic field is applied along the $c$ axis. Between the applied field 7T and 13T, there exists a noncollinear phase with a modulation vector  $(q, q, 0)$ with $q=\frac {1}{3}$.  At a  magnetic field above 13T, CuFeO$_2$ adopts the five-sublattice magnetic structure ($\uparrow\uparrow\uparrow\downarrow\downarrow$) with collinear moments along the $c$ axis.\cite{Petrenko2000,Mekata1993} An in-plane electric polarization is observed only in the intermediate-field (between 7T and 13T) when the  system adopts proper screw type of  magnetic ordering.  There are three possible mechanism for electric polarization in improper multiferroics: (i) magnetostriction, (ii) spin current model\cite{knb} or inverse Dzyaloshinskii-Moriya (DM) effect,  and (iii) spin-orbit coupling dependent $d$-$p$ hybridization. Due to strong coupling between magnetism and ferroelectricity, the improper multiferroelectrics are very interesting to study.\cite{nicolaScience, khomskii2006, ChakrabortyPRB13} Arima \cite{Arima2007} showed that  the electric polarization in CuFeO$_2$ can be explained  by the third mechanism, when the proper screw type of magnetic ordering can induce ferroelectricity through the variation in the metal (Fe-3$d$)-ligand (O-$p$) hybridization with spin-orbit coupling. Other delafossites CuCrO$_2$ and AgCrO$_2$ also  exhibit ferroelectricity for a particular kind  of magnetic ordering.\cite{Mekata1993, Soda2009} AgCrO$_2$  exhibits ferroelectric polarization below the temperature 21 K, in which each triangular layers of Cr$^{3+}$ ions form parallel chains with helical spiral spin order.\cite{Seki2008, Kan2009} 
The ferroelectric polarization is also observed for CuCrO$_2$  with a noncollinear 120$^\circ$ spin structure $\9q \sim (\frac{1}{3},\frac{1}{3},0)\0$, below 23.6 K. 
 
The triangular lattice antiferromagnet rhombohedral (3$R$) AgFeO$_2$ has recently been under focus  after the synthesis of the high quality samples under high pressure by Tsujimoto {\em et al.}\cite{Terada2012}  Silver ferrite crystallizes in hexagonal structure also. \cite{Okamoto-1972} 3$R$-AgFeO$_2$ shows interesting magnetic, thermodynamic and ferroelectric properties. \cite{Terada2012, Vasiliev2010, Terada2015}
3$R$- AgFeO$_2$ exhibits negative Curie-Weiss temperature ($\theta_{\text{CW}}=-140$ K) indicating  antiferromagnetic interactions between Fe$^{3+}$ (3$d^5$) ions.
The system has two magneto-structural phase transitions at $T_{N1}= 15$ K and  $T_{N2}= 9$ K. The magnetic structure (ICM1 phase) for $9$ K$\le T\le 15$ K is a spin-density wave with  incommensurate propagation vector $k= (1, q, \frac{1}{2})$ with $q=0.384$. At temperature below 9 K, the magnetic structure (ICM2 phase)  of 3$R$-AgFeO$_2$ turns into an elliptical cycloid with the incommensurate propagation vector $k= (-\frac{1}{2}, q, \frac{1}{2})$ with $q=0.2026$.\cite{Terada2012}. The magnetic ground state of AgFeO$_2$ is drastically different from the commensurate magnetic ground state of CuFeO$_2$, that indicates that the  A-site cation plays a crucial role in magnetism. In ICM2 phase of 3$R$-AgFeO$_2$, the system  shows ferroelectric polarization ($\sim$ 300 $\mu$C/m$^2$) for the powder sample.  It is suggested that polarization is possibly driven by the spin current mechanism.\cite{knb} In view of the above it is important to investigate the origin of ferroelectricity in AgFeO$_2$.  Since by replacing the nonmagnetic A-site (Cu with Ag)  the magnetic properties are drastically changed,  it is also interesting to compare the electronic and magnetic properties of these two systems.

In this paper, we have employed {\em ab~initio} density functional calculations to  investigate the electronic, magnetic and ferroelectric properties of the two dimensional system  AgFeO$_2$, where Fe$^{3+}$ ions (3$d^5$, $S=\frac {5}{2}$) form a triangular lattice and also  compare our results with the isostructural analogue CuFeO$_2$. The remainder of this paper is organised as follows: in section \ref{sec:crystal} we have described the crystal structure and computational details. Section \ref{sec:results} is devoted to the detailed discussion of our results on electronic structure calculations. Finally we conclude in section \ref{sec:concl}. 
\section{\label{sec:crystal} Crystal structure and Computational details}

\begin{figure}[h]
\centering
\includegraphics[scale=.43]{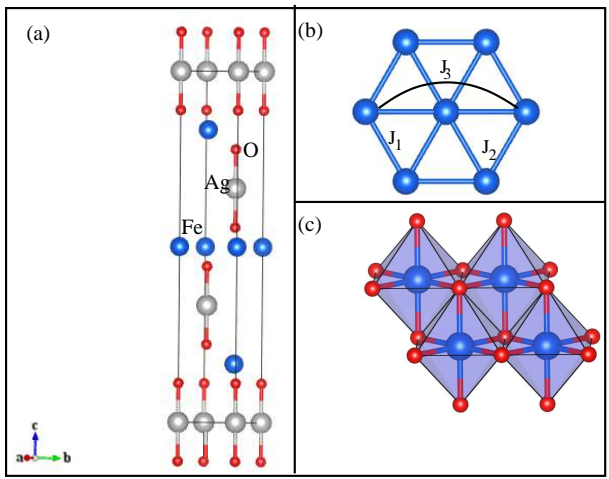}
\caption{\label{fig:structure} (a) The unit Cell of AgFeO$_2$, (b) Hexagonal $ab$ plane, (c) Edge sharing FeO$_6$  octahedra.}
\end{figure}

 The crystal structure of 3$R$-AFeO$_2$, (A $=$ Cu, Ag) shown in Fig.~\ref{fig:structure} belongs to the rhombohedral space group $R\bar{3}m$. The silver ferrite consists of triangular layers of slightly distorted edge-sharing FeO$_6$  octahedra. The Ag$^+$ ions are in between the FeO$_2$ plane and are in a dumbbell (O- Ag$^+$- O) coordination as shown in Fig.~\ref{fig:structure}.The structural informations of AgFeO$_2$ and CuFeO$_2$ are taken from Ref.~\citenum{Shannon}.

The density-functional-theory (DFT) calculations were carried out within two different methods: (a) the plane-wave-basis-based projector augmented wave (PAW) \cite{blochl} method  as implemented in the Vienna {\textit{ab~initio}} Simulation package (VASP)\cite{vasp}, (b) the linear-muffin-tin-orbital (LMTO) basis within atomic sphere  approximation (ASA) with Stuttgart TB-LMTO-ASA-47 code\cite{Anderson}. The basis set for the self-consistent electronic structure calculations for AgFeO$_2$ in TB-LMTO ASA includes Ag $(s, p, d)$, Fe $(s, p, d)$, and O $(s, p)$ and the rest are downfolded. The density of states calculated using the TB-LMTO ASA method is found to be in good agreement with the density of states calculated using plane wave basis.  
We have analyzed the chemical bonding by computing the crystal orbital Hamiltonian population (COHP) as implemented in the
Stuttgart tight-binding linear muffin-tin orbital (TB-LMTO) code\cite{Anderson}. The COHP provides the information regarding the specific pairs of atoms that participate in the bonding, and also the range of such interactions.
  
For plane wave based calculations, we  have used a plane-wave energy cutoff of 500 eV and $\Gamma$ centered $k$-space sampling on  ($4\times4\times1$) $k$-mesh.  The localized Fe-$d$ states are treated in the framework of LSDA+U method.\cite{dudarev1998}. In order to find out the importance of spin-orbit coupling (SOC) we have also carried out the electronic  structure calculation with SOC  in the framework of the LSDA+U+SOC method.  All structural relaxations are carried out until the Hellman-Feynman forces on each atom became less than 0.01  eV/\AA. To estimate the ferroelectric polarization we have used Berry phase method\cite{resta1994} as implemented in the Vienna {\textit{ab initio}} simulation package (VASP)\cite{vasp}.

\section{\label{sec:results}Results and discussions}
\subsection{Electronic and magnetic properties}
\subsubsection{Spin unpolarized calculation}
\begin{figure}
\centering
\includegraphics[scale=.25]{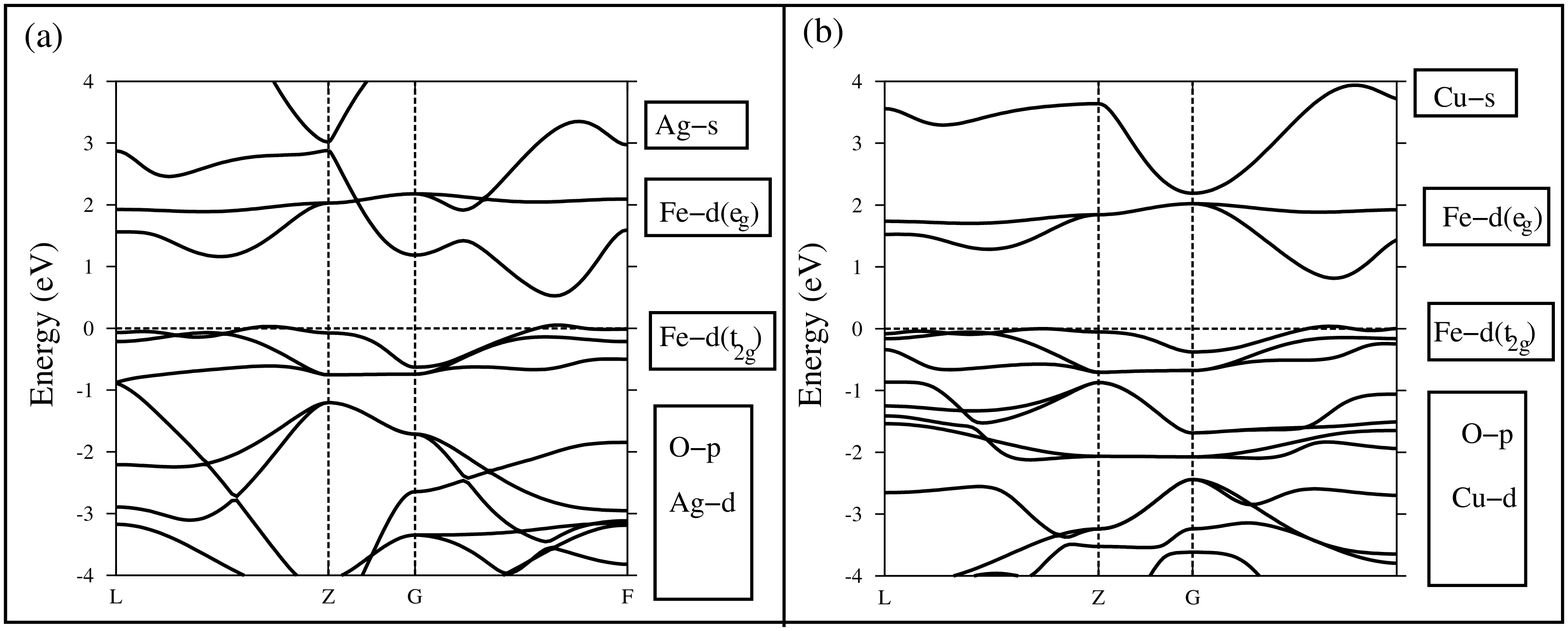}
\caption{\label{fig:bandstructure} LDA band structure of (a) AgFeO$_2$ and (b) CuFeO$_2$. The zero of the energy has been set up at the LDA Fermi energy.}
\end{figure}

To get insight on the electronic structure of AgFeO$_2$ and CuFeO$_2$, we have started with spin-unpolarized calculations. The band structures calculated with the LMTO method for these systems are plotted along the various high symmetry points of the Brillouin zone of the rhombohedral lattice (see Fig.~\ref{fig:bandstructure}).  The bands are plotted with respect to the Fermi energy $(E_F)$ of the compounds.  Since Fe is in Fe-$3d^5$ configuration and in an octahedral environment, the $t_{2g}$ states, that can accommodate 6 electrons, are partially occupied, while the $e_g$ bands are completely unoccupied.
 The Fe-$d$ bands are well separated from the filled  Ag-$d$ and O-$p$ bands for AgFeO$_2$ and Cu-$d$ and O-$p$ bands for CuFeO$_2$. 

\begin{figure}
\centering
\includegraphics[scale=.3]{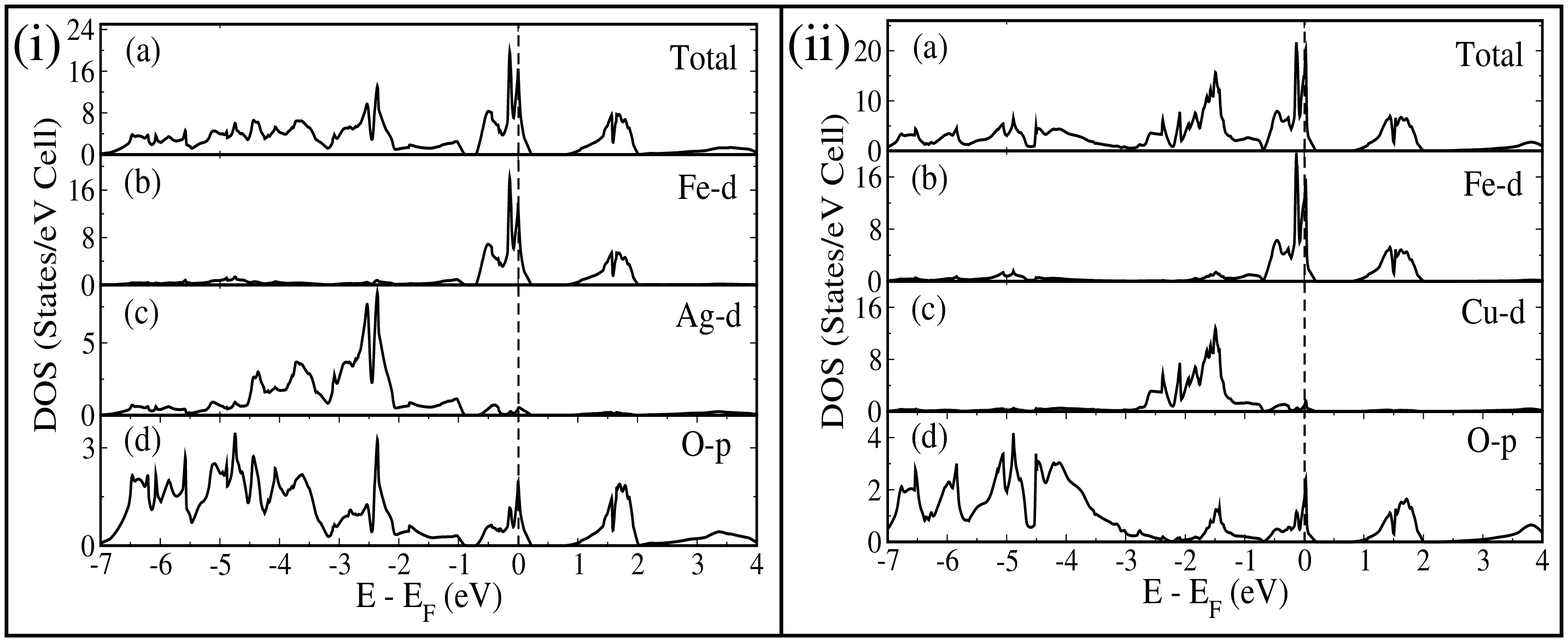}
\caption{\label{fig:nonmagdos} Total and partial density of states (i) for AgFeO$_2$ and (ii) for CuFeO$_2$.}
\end{figure}

 The total density of states (DOS) as well as its projection onto various atomic orbitals (PDOS) are shown in Fig.~\ref{fig:nonmagdos}. The DOS are projected onto Fe-$d$, Ag-$d$, and O-$p$ orbitals for  AgFeO$_2$ and Fe-$d$, Cu-$d$, and O-$p$ orbitals for  CuFeO$_2$ . The spin unpolarized calculation give rise to a metallic solution with states dominated by  Fe-$d$ character at the Fermi level $(E_F)$ for both the systems. The Fe-$d$  density of states are spread over $-$1 eV below the Fermi level to 2.0 eV above the Fermi level for both the systems (see Fig.~\ref{fig:nonmagdos}), and also hybridize with the O-$p$ states. For AgFeO$_2$, the Ag-4$d$ states are completely filled  and spread over $-1$ eV to $-7$ eV below the Fermi level (see Fig.~\ref{fig:nonmagdos}(i)(c)). On the other hand for CuFeO$_2$,  the Cu-3$d$ states are completely filled  and spread over $-0.5$ eV to $-3$ eV below the Fermi level (see Fig.~\ref{fig:nonmagdos}(ii)(c)). Both the Ag-$d$ and Cu-$d$ states also hybridize with oxygens. 
\begin{figure}
\centering
\includegraphics[scale=.35]{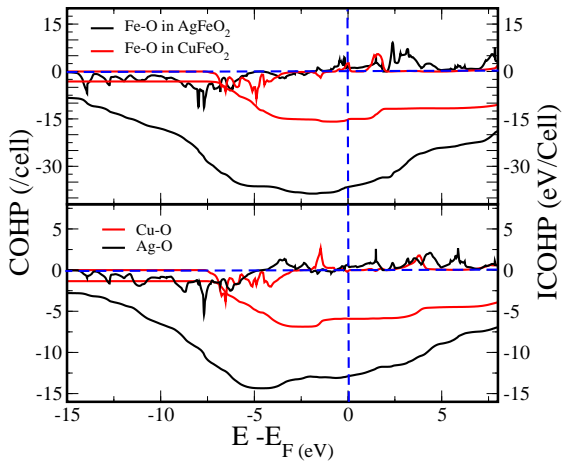}
\caption{\label{fig:cohp} (i) COHPs and integrated COHPs (ICOHP) per bond (a) for Fe-O for AgFeO$_2$ and CuFeO$_2$, and (b) Ag-O and Cu-O for  AgFeO$_2$ and CuFeO$_2$ respectively.}
\end{figure}

 We have also compared the hybridization of Fe-O and also A-O (A$=$ Cu, Ag) for these two systems by analyzing the COHP plots, that provide an energy resolved visualization of the chemical bonding. In COHP, the density of states is weighted by the Hamiltonian matrix elements where the off-site COHP represents the covalent contribution to bands. The bonding contribution for which the system undergoes a lowering in energy is indicated by negative COHP and the antibonding contribution that raises the energy is represented by positive COHP. Thus it gives a  quantitative measure of bonding. In Fig.~\ref{fig:cohp} we have plotted the off-site COHP and the energy integrated COHP (ICOHP) per bond for the nearest neighbor Fe-O and Ag-O for AgFeO$_2$ and Fe-O and Cu-O for CuFeO$_2$.  From the COHP plots in Fig.~\ref{fig:cohp}, we find that strongest covalency is between Fe and O for both the systems. The Ag-O covalency is substantially stronger in comparison to Cu-O covalency as revealed by the integrated COHP at the Fermi level where ICOHP value for Ag-O is $-$6.44 eV and for  Cu-O, ICOHP value is $-$2.95 eV.  Interestingly, the nearest-neighbor Fe-O covalency is larger for AgFeO$_2$ (ICOHP values are $-$6.092 and $-$2.53 eV for  AgFeO$_2$ and CuFeO$_2$ respectively). This difference in hybridization with oxygen  may be responsible for different magnetic ground state of  AgFeO$_2$  and CuFeO$_2$. 

\subsubsection{Spin polarized calculations} 
\begin{figure}
\centering
\includegraphics[scale=.4]{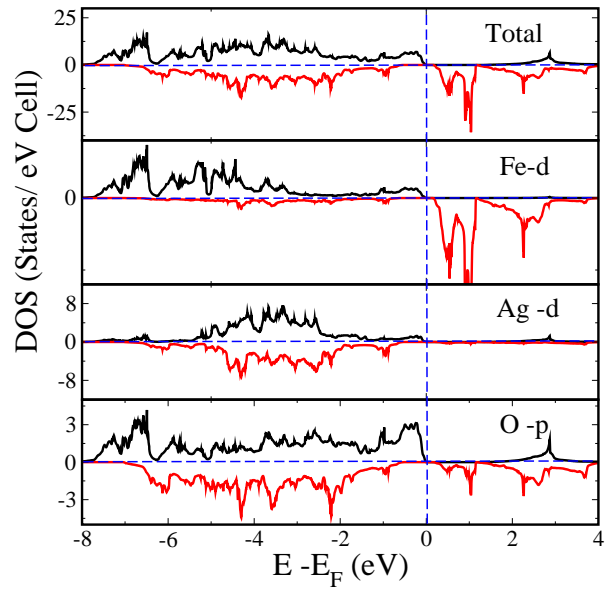}
\caption{\label{fig:fmdos} Total and orbital decomposed density of states for  AgFeO$_2$.}
\end{figure}

In order to study the magnetic properties we have carried out  spin-polarized calculations for AgFeO$_2$. The total and orbital decomposed density of states for ferromagnetic configurations are plotted in Fig.~\ref{fig:fmdos}. As suggested in Ref.~\onlinecite{Ong2007}, we have used the onsite Coulomb interaction term $U=3$ eV and the onsite exchange interaction $J=1$ eV  for Fe $d$ states in  AgFeO$_2$. From Fig.~\ref{fig:fmdos}(a), we find that the ferromagnetic state is insulating with $U=3$ eV and $J=1$ eV in LSDA+U calculation. The plot of the  density of states (DOS) reveal the presence of five electrons in  Fe-3$d$  spin up channel, which is consistent with the picture of high-spin Fe$^{3+}$  ions in AgFeO$_2$. The O-$p$ and Ag-$d$ states are completely occupied. In ferromagnetic configuration, the spin moment of Fe site is  4.23 $\mu_B$ with $U=3$ eV and $J=1$ eV, and the rest are partly accommodated in O ($m_{\text{O}}=0.09 \mu_B$).
\subsubsection{Symmetric exchange interactions}
\begin{figure}[h]
\centering
\includegraphics[scale=.4]{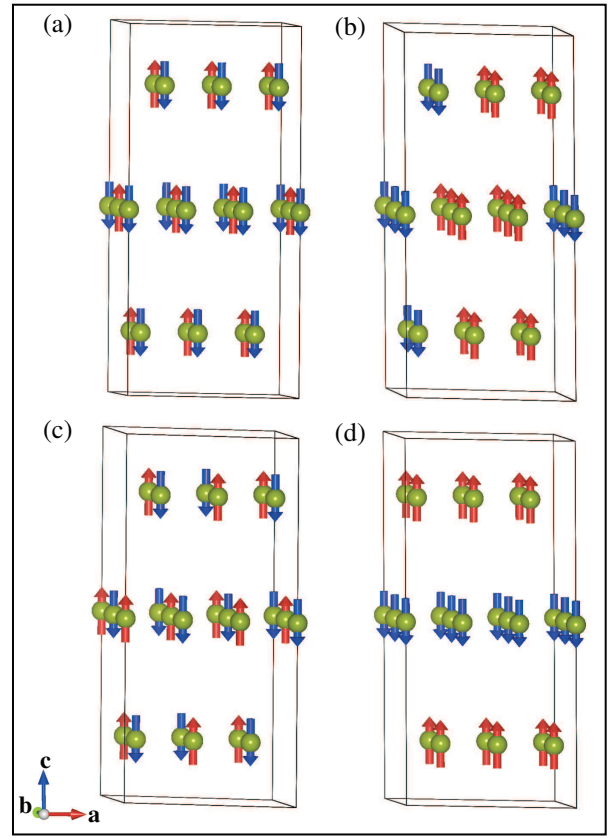}
\caption{\label{fig:exchange}  The four ordered spin states (AF1, AF2, AF3, AF4) of AgFeO$_2$ constructed using a $\9 3\times2\times1\0$ supercell. (a) AF1: spins are arranged ferromagnetically along $a$ and $c$ direction and antiferromagnetically along $b$ direction, (b) AF2: the spin arrangement is $\uparrow$$\downarrow$$\downarrow$  along  $a$ direction   and ferromagnetic along other two directions, (c) AF3: spins are antiferromagnetically arranged along $a$ and $b$ directions and ferromagnetically along $c$ direction, (d) AF4: spin arrangement is such that intra-layer couplings are ferromagnetic and inter-layer couplings are antiferromagnetic.}
\end{figure}

In order to determine the various exchange parameters ($J_{i}$), (indicated in Fig.~\ref{fig:structure}(b)), we have first calculated the total energies of several ordered spin states of a system and then related the energy differences between these states to the corresponding energy differences expected from the Heisenberg spin Hamiltonian:\cite{XiangPRB2007, JayitaPRB2012, kotibsco}
\begin{eqnarray}
H &=& \sum_{i,j}{\large J_{ij}}{\vec {S}_i}\cdot{\vec {S}_j}
\end{eqnarray}
We have calculated the total energies of five ordered spin states (FM, AF1, AF2, AF3, AF4) shown in Fig.~\ref{fig:exchange}. The total spin exchange energies (per three formula units) of the five ordered spin states are expressed as:
\begin{eqnarray}
E(FM) &=& \frac{25}{4}(-9J_1  -9J_2 -9 J_3 -9J_4)\nonumber   \\
E(AF1) &=& \frac{25}{4}(3J_1 + 3J_2 -9J_3 -J_4) \nonumber \\
E(AF2) &=& \frac{25}{4}(-J_1 +3J_2 -J_3 -J_4) \nonumber\\
E(AF3) &=& \frac{25}{4}(3J_1 - J_2 - J_3 -J_4) \nonumber\\
E(AF4) &=& \frac{25}{4}(-9J_1 -9J_2 -9J_3 + 3J_4)
\end{eqnarray}

The relative energies for AgFeO$_2$, calculated from LSDA+U method are summarized in Table \ref{tab:energy1}. 
The  nearest neighbor,  next-nearest-neighbor and diagonal interactions are denoted by $J_1$,  $J_2$  and $J_3$ respectively. $J_4$ denotes the coupling between nearest neighbors in adjacent layers. The exchange interactions are displayed in Table \ref{tab:exchange}. In last column of Table \ref{tab:exchange}, the values in the parentheses are  the exchange interactions  for CuFeO$_2$ obtained from Ref.~\citenum{Zhang2011}.
\begin{table}
\begin{center}
\caption{\label{tab:energy1} Relative energies per three f.u (in meV) for AgFeO$_2$ determined from LSDA+U calculations.}
\begin{ruledtabular}
\begin{tabular}{c..}
Configuration  & \multicolumn{1}{c}{$\Delta E$} & \multicolumn{1}{c}{$\Delta E$}\\
& \multicolumn{1}{c}{(for $U_{eff}=2$ eV)} & \multicolumn{1}{c}{(for $U_{eff}=4$ eV)} \\
\hline
FM & 0.0 & 0.0\\
AF1  & -234.5 & -99.54\\
AF2  & -212.33 & -94.03\\
AF3  & -254.5 & -111.57\\
AF4 & -66.0 & -31.4  \\
\end{tabular}
\end{ruledtabular}
\end{center}
\end{table}
\begin{table}
\begin{center}
\caption{\label{tab:exchange} Symmetric exchange interactions (in meV) for  AgFeO$_2$ and CuFeO$_2$ are tabulated here. The exchange interactions for CuFeO$_2$ inside the parentheses,  are adapted from Ref.~\citenum{Zhang2011}. }
\begin{ruledtabular}
\begin{tabular}{....}
\multicolumn{1}{c}{Exchange}  &   \multicolumn{2}{c}{AgFeO$_2$} & \multicolumn{1}{c}{CuFeO$_2$}\\
&  \multicolumn{1}{c}{$U_{eff}=2$ eV}  & \multicolumn{1}{c}{$U_{eff}=4$ eV} & \multicolumn{1}{c}{$U_{eff}=4$ eV} \\
\hline
J_1 & -1.92 & -0.83 & -0.77 (-0.76)\\
J_2 & -0.62  & -0.21 & -0.15 (-.18)\\
J_3 & -0.81 &  -0.35 & -0.28 (-.30)\\
J_4 & -0.88 & -0.42 &-0.24 (-.23)\\
\end{tabular}
\end{ruledtabular}
\end{center}
\end{table}
We find that all the exchange interactions ($J_1$, $J_2$, $J_3$, and $J_4$) are antiferromagnetic type. 
As a consequence both the intra-layer and the inter-layer exchange interactions are spin-frustrated in AgFeO$_2$ as well as  in CuFeO$_2$. In AgFeO$_2$,  the inter-layer exchange interaction $J_4$ is strongest in comparison to intra-layer exchange interactions $J_2$ and    $J_3$, while $J_3$ is dominant in CuFeO$_2$.
The exchange interactions are of super exchange type for both the systems. The intra-plane $J_1$ exchange interaction is mediated by Fe-O-Fe super-exchange path. Other intra-plane exchange interactions $J_2$ and  $J_3$ are mediated through Fe-O$..$O-Fe super super exchange path. The inter planer exchange coupling $J_4$ is mediated via the Fe-O-A-O-Fe path. As we have shown by plotting COHP (see Fig.~\ref{fig:cohp}) that the hybridization of Ag with oxygen for AgFeO$_2$ is greater than the hybridization of Cu with oxygen for CuFeO$_2$, the inter layer coupling is much stronger for AgFeO$_2$.   The antiferromagnetic nature of the exchange interactions in each FeO$_2$ layer results in spin frustration in the $\9 J_1, J_1, J_1 \0$ and $\9 J_2, J_2, J_2 \0$ triangles and in the $\9 J_1, J_1, J_2 \0$ line segments. Between adjacent FeO$_2$ layers, spin frustration occurs in the isosceles $\9 J_1, J_4, J_4 \0$  triangles.  The silver ferrite appears to be a more frustrated system with higher values of exchange interaction parameters as compared to the copper ferrite.
We have also calculated the  Curie-Weiss temperature $\theta_{\text{CW}}$ for AgFeO$_2$. In the mean field limit, the Curie-Weiss temperature $\theta$ is related to the exchange interactions as follows:
\begin{eqnarray}
\theta &=& \frac{S(S+1)}{3K_B} \sum_i{z_iJ_i}
\end{eqnarray}
where, the summation runs over all nearest neighbors of a given spin site, $z_i$ is the number of nearest neighbors connected by the spin exchange interaction $J_i$, and $S$ is the spin quantum number of each spin site (i.e., $S= 5/2$ in the present case). The calculated Curie-Weiss temperature ($\theta_{\text{CW}}$) for AgFeO$_2$ is $-363$ K and $-252$ K for $U_{eff}=$2 and 4 eV respectively, while the experimental value is $-140$ K. Thus, according to the experimental Curie-Weiss temperature and the mean-field theory, the calculated $J_1$-$J_4$ values are overestimated by a factor of $f =$ 2.6 and 1.9 for $U_{eff}=$2 and 4 eV respectively. The  $\theta_{\text{CW}}$ for CuFeO$_2$ is  $-292$ K  with   overestimation factor $f = 3.24$ as reported in Ref.~\citenum{Zhang2011}.  From the magnitude of the symmetric exchange interactions for AgFeO$_2$ and  CuFeO$_2$, we conclude that the difference in the magnetic ground state for AgFeO$_2$ and CuFeO$_2$ probably do not stem from the symmetric part of the spin Hamiltonian. In the following, we have investigated the magnetic properties of AgFeO$_2$ and CuFeO$_2$ including spin orbit interaction.

\subsubsection{Spin-orbit coupling}
The importance of spin-orbit coupling in triangular lattice antiferromagnets has been discussed in literature. \cite{TanakaPRL2012, Terada2012} In this work we have investigated the importance of spin-orbit coupling and single ion anisotropy for AgFeO$_2$ and CuFeO$_2$.
 We have considered the FM spin configurations for both the system and included the spin orbit coupling (SOC) in the framework of LSDA+U+SOC calculations. The orbital moment at the Fe site is $\sim$ 0.028  $\mu_B$ and $\sim$ 0.025 $\mu_B$ for AgFeO$_2$ and CuFeO$_2$ respectively. For high spin configuration of Fe$^{3+}$ (3$d^5$), the orbital moment is expected to be quenched. Here induced mechanism due to either mixing of Fe-$d$ with oxygen $p$ states or $t_{2g}$-$e_g$ orbitals is possible leads to finite orbital moment. In fact, our COHP analysis suggests substantial hybridization between Fe and oxygen. We have calculated the total energy by choosing the various spin quantization axes, and the result of our calculation is displayed in Table~\ref{tab:MCA}.\cite{Chakraborty2016,Jayita-JPCM2017} An estimation of  magnetocrystalline anisotropy is obtained from the energy difference between calculations with spin quantization chosen along the $c$ direction $\9 0  0  1\0$ and perpendicular to the $c$ direction, yield values 0.33 meV and 0.21 meV per Fe ion for AgFeO$_2$ and  CuFeO$_2$ respectively within LSDA+U+SOC (for $U_{eff}=4$ eV) calculation. The magnetocrystalline anisotropy energy is larger for AgFeO$_2$ than CuFeO$_2$ indicating important role of SOC for AgFeO$_2$.


\begin{table*}
\begin{center}
\caption{\label{tab:MCA} The energy differences  between calculations with spin quantization chosen along different directions within  LSDA+U+SOC calculations.}
\begin{ruledtabular}
\begin{tabular}{c....}
Quantized axis  & \multicolumn{2}{c}{$\Delta E$ for AgFeO$_2$} & \multicolumn{2}{c}{$\Delta E$ for CuFeO$_2$} \\
& \multicolumn{1}{c}{$U_{eff}=2$ eV} & \multicolumn{1}{c}{$U_{eff}=4$ eV} & \multicolumn{1}{c}{$U_{eff}=2$ eV} & \multicolumn{1}{c}{$U_{eff}=4$ eV}\\
\hline
$(0 0 1)$ & 0.0 & 0.0  & 0.0 & 0.0 \\
$(0 1 0)$ &  0.83 & 0.33 & 0.65 & 0.21 \\
$( 1 0 0 )$ &  0.83 & 0.33 & 0.65 & 0.21 \\
\end{tabular}
\end{ruledtabular}
\end{center}
\end{table*}

\subsubsection{Antisymmetric exchange interactions}
Next, we have  considered the antisymmetric part of the spin Hamiltonian $H = \sum_{\braket{ij}} \vec{D}_{ij} \cdot \9\vec{S}_i \times \vec{S}_j\0$ and calculated the Dzyaloshinskii-Moriya interactions parameter ($\vec{D}$) from the total energy calculations with spin orbit coupling as discussed in Ref.~\citenum{Xiang2011}. Here we have  calculated the three components $D^x_{12}$, $D^y_{12}$, $D^z_{12}$ of the DM vector (between nearest neighbor spin sites 1 and 2) for AgFeO$_2$ and CuFeO$_2$ by performing LSDA + U + SOC calculations. In order to calculate $x$ component of $\vec{D}_{12}$, we consider the following four spin configurations in which the spins 1 and 2 are oriented along the $y$ and $z$ axes, respectively: (i) $S1 = (0,S,0)$, $S2 = (0,0, S)$, (ii) $S1 = (0, -S, 0)$, $S2 = (0, 0, S)$, (iii) $S1 = (0, S, 0)$, $S2 = (0, 0, -S)$, (iv) $S1 = (0, -S, 0)$, $S2 = (0, 0, -S)$. In these four spin configurations, the spins of all the other spin sites are the same and are along the $x$ direction. The spin interaction energy for the four spin configurations can be written as
\begin{eqnarray}
E_{spin}= E_{other}+ D^x_{12}S^y_1S^z_2-S^y_1\sum_{i=3,4}D^z_{1i}S^x_i+ S^z_2\sum_{i=3,4}D^y_{2i}S^x_i
\end{eqnarray}
Similarly for $y$ and $z$ components of $\vec{D}_{12}$.
Our calculated values (in meV) of the components of  $\vec{D}_{12}$ for AgFeO$_2$ are $D^x_{12}=0.0104$, $D^y_{12}=-0.42$ and  $D^z_{12}=0.005$ and the magnitude of DM vector is  0.4202 ($\frac{|D_{12}|}{J_1} =$ 0.55 ) for $U_{eff}=4$ eV. We have also computed $\vec{D}_{12}$  for $U_{eff}=2$ eV and corresponding components of $\vec{D}_{12}$ are $D^x_{12}=0.011$, $D^y_{12}=-0.87$, $D^z_{12}=0.01$ and $\frac{|D_{12}|}{J_1} =$ 0.5. The large $y$ component of $\vec{D}_{12}$  term makes the two spins perpendicular to each other in the $ac$-plane.\cite{Lu2012,Li2012}.

We have also calculated the components of $\vec{D}_{12}$ (between nearest neighbor spin sites 1 and 2)  for CuFeO$_2$. The components are $D^x_{12}=0.004$, $D^y_{12}=-0.16$,  $D^z_{12}=0.012$ and the magnitude of DM vector is  0.161 ($\frac{|D_{12}|}{J_1} =$ 0.2) for $U_{eff}=4$ eV. For $U_{eff}=2$ eV, the components are $D^x_{12}=0.004$, $D^y_{12}=-0.23$, and  $D^z_{12}=0.012$ with $\frac{|D_{12}|}{J_1} =$ 0.15.  The DM vector for nearest neighbor sites,  is much smaller for CuFeO$_2$ compared with AgFeO$_2$.
These calculations suggest that SOC has a profound impact on AgFeO$_2$ and plays a key role in determining its magnetic ground state.  

\subsubsection{Low temperature magnetic structure}
\begin{figure}[h]
\centering
\includegraphics[scale=.35]{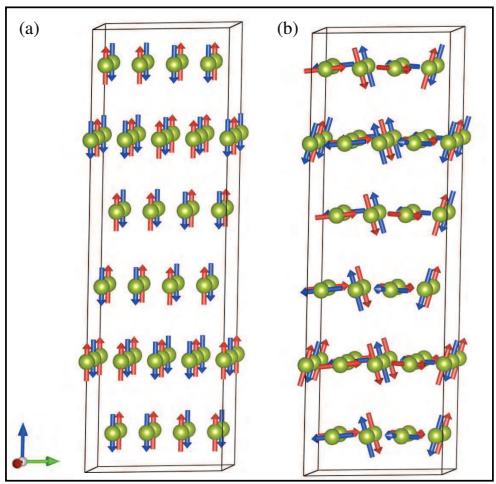}
\caption{\label{fig:magconfig} (a) AFM1 collinear magnetic ordering (b) AMF2 noncollinear magnetic ordering.}
\end{figure} 
In order to simulate the low temperature (below 9 K)  magnetic order of AgFeO$_2$, we have made a (2$\times$4$\times$2) supercell  which contains 192 atoms.  We have considered two magnetic configurations AFM1 and AFM2 as shown in Fig.~\ref{fig:magconfig}. In AFM1 configuration $(q=(\frac{1}{2}, \frac{1}{4}, \frac{1}{2}))$ the spin arrangement along $b$ direction is  $\uparrow \uparrow \downarrow \downarrow $  and they are antiferromagnetically aligned along $a$ and $c$ direction.  For AFM2 configuration $\9q=(\frac{1}{2}, \frac{1}{4}, \frac{1}{2})\0$, we have made the noncollinear spin arrangements along $b$ directions. Spins are antiferromagnetically aligned along $a$ and $c$ directions. The three components of spin at Fe sites for a Fe-O layer is displayed in Table ~\ref{tab:spincomp}.
 \begin{table}
\begin{center}
\caption{\label{tab:spincomp}The three  components of spin (in $\mu_B$) for one Fe-O layer in AFM2 magnetic configuration for AgFeO$_2$ determined from LSDA+U calculations ($U_{eff}=2$ eV).}
\begin{ruledtabular}
\begin{tabular}{...}
m_x  & m_y & m_z \\
\hline
-0.822  & 2.690  &  3.090\\
 0.801 & -2.600  &  -3.152 \\ 
 3.038 &  2.872  &  0.012\\
 0.128  &  2.801 & -3.100\\
 -3.027 &  2.875 & -0.215\\
 -0.165 & -0.363 &  4.157 \\
 -3.046 & -2.845 & -0.228\\
  3.055 & -2.845 &  0.019\\
\end{tabular}
\end{ruledtabular}
\end{center}
\end{table}

\begin{figure}
\centering
\includegraphics[scale=.3]{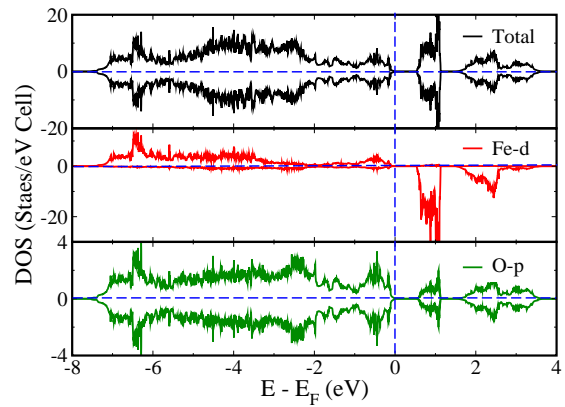}
\caption{\label{fig:afmdos} The plot of total and orbital projected density of states in AFM1 magnetic configuration for  AgFeO$_2$ within LSDA+U calculation (for $U_{eff}=$2  eV).}
\end{figure}

We have calculated the electronic structure for the FM, AFM1, and AFM2 using LSDA+U method. The results of our calculations are displayed in Table~\ref{tab:EnergyDifference} and we find that the AFM1 magnetic configuration has the lowest in energy. The total and partial density of states in the AFM1 magnetic configuration for  AgFeO$_2$ calculated using LSDA+U method, is shown in Fig.~\ref{fig:afmdos}. The band gap in AFM1 state is calculated to be 0.54 eV and 1.03 eV for $U_{eff}=$2 and 4 eV respectively. The majority Fe-$d$ states are completely occupied while the minority states are empty, which is consistent with the Fe$^{3+}$ valence state of Fe with a $3d^5$ configuration. Such a half-filled configuration promotes the antiferromagnetic order. The spin magnetic moment of Fe is 3.982 $\mu_B$ and 4.18  $\mu_B$ with $U_{eff}=$2  and 4 eV respectively. 

We have added the  spin orbit coupling (SOC) in our calculation. The results of our calculation for AFM1 and AFM2 structures is displayed in Table~\ref{tab:EnergyDifference}. Our calculations clearly reveal AFM2 is the ground state for AgFeO$_2$ upon inclusion of SOC, indicating important role of spin orbit interaction as anticipated earlier. The spin and orbital moments at Fe sites are 4.15 $\mu_B$ and 0.027 $\mu_B$ respectively.  
\begin{table}
\begin{center}
\caption{\label{tab:EnergyDifference} The relative energies  (in meV) for AFM1,  AFM2, and FM  magnetic configurations within LSDA+U and LSDA+U+SOC calculations.}
\begin{ruledtabular}
\begin{tabular}{ccc}
Configurations & LSDA+U &  LSDA+U+SOC\\
AFM1 & -24  & 76\\
AFM2 & 0.0  & 0.0\\
FM & 43  & 86\\
\end{tabular}
\end{ruledtabular}
\end{center}
\end{table}
\subsection{Ferroelectric properties}
Finally, we have calculated the ferroelectric polarization with AFM1 and AFM2 magnetic configuration  using Berry phase method\cite{resta1994} as  implemented in the VASP\cite{vasp}. We do not find the electric polarization for the centrosymmetric crystal structure with AFM1 and AFM2 magnetic structure within LSDA+U for $U_{eff}=2$ eV.   With the application of spin orbit coupling (SOC), there is no polarization for AFM1 structure,  however  AFM2 magnetic configuration attains a polarization value of 34 $\mu$$C/m^2$ respectively, which suggests that the noncollinear magnetic order induces polarization via spin orbit coupling.  Therefore spin-current mechanism is one  source of the electric polarization in this system. 
But our calculated  magnitude of electric polarization is much smaller than the observed experimental value of polarization ($\sim$300 $\mu$$C/m^2$ for powder sample). This result suggests that although  there is electronic contribution to the polarization but the lattice mechanism also important for the ferroelectric polarization in AgFeO$_2$.
\begin{table}
\begin{center}
\caption{\label{tab:distancerela} The bond length  between the magnetic atoms in experimental structure and change in bond length in relaxed structures. (change in bond length $-$Ve means the shortening of bond length after relaxation)}
\begin{ruledtabular}
\begin{tabular}{cccc}
Exchange  & Distance  (\AA) & change in bond  &  change in bond   \\
paths & Exp. Struc. & length (\AA)  &   length (\AA) \\
& & (AFM1 relax) & (AFM2 relax)\\
\hline
$J_1$  & 3.09 & $-$0.04 & $-$0.09\\
$J_2$ & 5.26 & 0.01 & 0.06\\
$J_3$  & 6.08 &  $-$0.01 & 0.02\\
$J_4$ &  6.44 & 0.0 & 0.01\\
\end{tabular}
\end{ruledtabular}
\end{center}
\end{table}

In order to obtain the lattice contribution to polarization, we have carried out relaxation calculations within the framework of LSDA+U and LSDA+U+SOC method.  In this optimization, the cell parameters were fixed to the experimental values, but the positions of the atoms were allowed to relax. The maximum change in bond lengths occurs for the nearest neighbor. The bond lengths before and after relaxations are listed in Table~\ref{tab:distancerela}.  
The ferroelectric polarization of  282 $\mu$$C/m^2$ is found for the relaxed  AFM1 collinear structure without SOC. This result suggests that symmetric spin exchange induces the polarization in this material through the exchange-striction mechanism. Application of SOC, the polarization turn out to be 324 $\mu$$C/m^2$.  The polarization is increased for relaxed AFM2 noncollinear structure, and the value of polarization is 407  $\mu$$C/m^2$  with $U_{eff}=$2 eV in LSDA+U calculation without SOC. We find that the polarization in relaxed structure with AFM2 magnetic configuration with spin-orbit coupling is 485 $\mu$$C/m^2$. We have also calculated the electric polarization with a larger value of $U$. Our results are summarized in Table~\ref{tab:Polarization}.  Our results suggest that the exchange striction mechanism as well as spin current model\cite{knb} are responsible for electric polarization in this system.

\begin{table}
 \begin{center}
\caption{\label{tab:Polarization} Calculated Polarization $(\mu$$C/m^2)$ in various magnetic structures}
\begin{ruledtabular}
\begin{tabular}{c..}
Structure & \multicolumn{1}{c}{P} &\multicolumn{1}{c}{P} \\
& \multicolumn{1}{c}($U_{eff}=$2 eV) &  \multicolumn{1}{c}($U_{eff}=$4 eV)\\
\hline
Exp. AFM1  & 0  & 0\\
Exp. AFM2  & 0  & 0\\
Exp. AFM2 + SOC &  34 & 26 \\
Relax. AFM1  & 282 & 264 \\
Relax. AFM1 + SOC & 324 & 308 \\
Relax. AFM2 &  407 & 373\\ 
Relax. AFM2 + SOC & 485 &  453\\
\end{tabular}
\end{ruledtabular}
\end{center}
\end{table}

\section{\label{sec:concl}Conclusion}
In this paper,  we have studied the electronic structure, magnetism, and ferroelectric properties of the triangular lattice antiferromagnet AgFeO$_2$ and compared our results with isostructural compound CuFeO$_2$. While Fe is in $d^5$ configuration in both the systems, the magnetic ground state of AgFeO$_2$ markedly different from CuFeO$_2$. In order to understand the origin of this difference, we calculated the symmetric exchange interactions offered by the Heisenberg model. Our calculations reveal that the symmetric exchange interactions are nearly identical for both the systems and therefore hardly play any role for the different magnetic ground state. Next, we incorporated spin-orbit coupling and our calculations regarding the orbital moment, magneto-crystalline anisotropy and DM parameters clearly indicate that SOC has a profound effect on AgFeO$_2$. It is interesting to note that SOC is operative in Fe 3$d^5$ manifold possibly by induced mechanism due to either mixing of Fe-$d$ with oxygen $p$ states or mixing of $t_{2g}$ - $e_g$ orbitals in the distorted octahedra. We recover the experimental magnetic ground state of  AgFeO$_2$  upon the inclusion of SOC. Calculations of ferroelectric polarization suggest that the spontaneous polarization arises from noncollinear spin arrangement via spin-orbit coupling.  Our calculations also indicate that in addition to electronic contributions, lattice mediated contribution to the polarization is also important for AgFeO$_2$.


%
\end{document}